\documentclass[aps,twocolumn,floatfix]{revtex4}
\usepackage{graphicx}
\usepackage{dcolumn}
\usepackage{bm}

\begin{document}
\title{Impurity bands and the character of the electronic states in ferromagnetic GaMnAs layers}

\author{E. Dias Cabral}
\affiliation{Instituto de F\'\i sica, Universidade do Estado do
Rio de Janeiro, Rio de Janeiro, R.J., Brazil}

\author{M. A. Boselli}
\affiliation{Departamento de F\'\i sica, Universidade Federal de
Ouro Preto,  Ouro Preto, M.G., Brazil}

\author{A. T. da Cunha Lima}
\affiliation{Universidade Veiga de Almeida, Cabo Frio, RJ, Brazil}

\author{I. C. da Cunha Lima}
\affiliation{Instituto de F\'\i sica, Universidade do Estado do
Rio de Janeiro, Rio de Janeiro, R.J., Brazil}

\date{\today}


\begin{abstract}
The interplay between disorder and spin polarization in a GaMnAs
thin layer results into spin-polarized impurity hole bands. A
figure of merit is defined to label the nature of the sample as
metallic or non-metallic. It is shown that samples with the
highest figures of merit have a ratio between the extended hole
density and the Mn concentration near 0.2, in agreement with the
ratio of 0.1-0.25 known to occur among samples produced with the
highest Curie temperatures. Both the non-metal-to-metal and the
metal-to-non-metal transitions experimentally observed in the
ferromagnetic regime are obtained, as the Mn concentration
increases.
\end{abstract}

\maketitle

Homogeneous samples of Ga$_{1-x}$Mn$_x$As were produced more than
ten years ago by Matsukura \textit{et al} \cite{matsu1,matsu2}.
Among these samples ferromagnetism has been observed in the range
$x=0.02-0.07$. The highest Curie temperatures were obtained near
$x=0.05$, and after annealing they approach 170K. In thin films
and trilayer systems Curie temperatures of 160 K were obtained for
$x=0.074$ \cite{newohno}. Samples with this high Curie temperature
have free hole concentrations near $1-2\times 10^{20}$ cm$^{-3}$,
obtained \textit{via} Hall measurements. This is a fraction
(10-25$\%$) of the total concentration of Mn, what is usually
explained as being due to the presence of As anti-sites and
interstitial Mn. At small Mn concentrations the alloy was observed
to be a paramagnetic insulator. As $x$ increases the alloy becomes
ferromagnetic, going through a non-metal-to-metal transition for
concentrations near 3\%. The transition occurs, then, at  very
high impurity concentrations, of the order of 1$\times 10^{20}$
cm$^{-3}$, as compared to transitions associated to shallow donor
occurring in Si and GaAs, near 1$\times 10^{18}$ cm$^{-3}$. For
$x$ above $5\%$, in GaMnAs, the alloy becomes, again, a
non-metallic ferromagnet \cite{matsu2}. The highest Curie
temperatures were found for the highest hole concentrations
\cite{matsu1}. Ferromagnetism in quasi-two-dimensional GaMnAs
layers  have been studied for their importance on having a spin
polarizing layer inside a semiconductor nanostructure
\cite{sado,us1}.

A survey on the optical and transport properties of GaMnAs samples
over a wide range of concentrations \cite{nottingham,prague}
identified the existence of an impurity band and explored the
character of the states near the Fermi level \cite{manyauthors}.
Impurity band in GaMnAs has been calculated \cite{usapl,fiete}. It
is known that the character of holes determines the indirect
interaction \cite{bhat1,nolting} among the magnetic ions,
influencing the Curie temperature of magnetically ordered samples.

In this work we address the entanglement of the magnetic and
transport properties of GaMnAs. We model a thin Ga$_{1-x}$Mn$_x$As
layer by a quasi-two-dimensional heavy hole gas of areal density
$n_s$ submitted to Coulomb scattering by a negative ionized
impurity system of concentration $n_i$. The two concentrations
$n_s$ and $n_i$ are considered as independent parameters. This is
important in the present context, since it allows for changing
$n_i$ and $n_s$ by co-doping \cite{furdyna2003} or controlling
$n_s$ by external means \cite{chiba2006,furdyna2007}. In addition
to the Coulomb scattering, the spins of the carriers are assumed
to interact with the localized magnetic moments at the Mn ions
through a Hubbard-like sp-d potential \cite{us1}.

At low temperatures (as compared to the Curie temperature) we can
neglect the  magnetic moment fluctuations on the impurities, which
otherwise produce spin-flip scattering. Assuming an homogeneous
distribution of the localized magnetic moments resulting into an
average normalized magnetization, $0\leq \langle M \rangle \leq
1$, the effective magnetic potential becomes $V_{mag}=-xN_0\beta
\langle M \rangle \sigma/2$. The hole spins are aligned
($\sigma=1$) or anti-aligned to the average magnetization
($\sigma=-1$). $x$ is the Mn doping factor, and $N_0\beta$ is the
exchange potential for holes, $N_0\beta=-1.2$eV \cite{matsu1}. If
$x=0.05$, $V_{mag}$ introduces a splitting of 150 meV for fully
magnetic ordered samples.

\begin{figure}
\includegraphics[angle=-90,width=8cm]{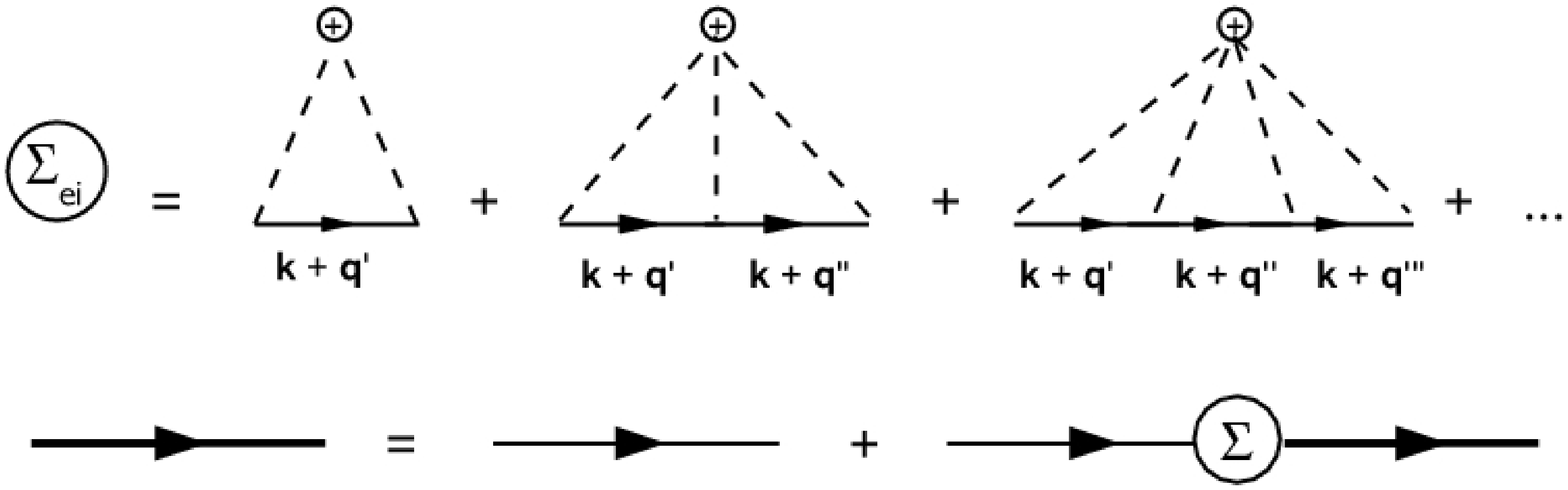}
\vspace{-2cm} \caption{Self-energy diagrams, within the multiple
scattering approximation. }\label{fig_a1}
\end{figure}

For a very large number of impurities, we can define an
``impurity-averaged'' Green's function,
$\overline{G(\textbf{q})}$, assuming they are randomly located
without any correlation. The multiple scattering approximation
consists in selecting from the self-energy insertions only those
consisting in the scattering which occur several times by the same
impurity, as described  in Figure \ref{fig_a1}.

It is convenient to define a vertex function \cite{ghaz1,usapl}:
\begin{eqnarray}
K(\textbf{k},\textbf{q}_1;E)&=&\frac{1}{(2\pi)^2}\int
d^2\textbf{q}'_1v(\textbf{q}_1-\textbf{q}'_1)\overline{G(\textbf{q}'_1)}\times
\nonumber
\\
& &[Nv(\textbf{k}-\textbf{q}'_1)+K(\textbf{k},\textbf{q}'_1;E)],
\label{voila}
\end{eqnarray}
where $N$ is the concentration of scatterers and $v(q)$ is the
Coulomb scattering potential. In terms of this vertex function,
the impurity self-energy becomes
$\Sigma_{\textrm{ei}}(\textbf{k},E)=K(\textbf{k},\textbf{k};E)$. A
function $U$ is defined as $ U(\textbf{k},\textbf{q};E)\equiv
K(\textbf{k},\textbf{q};E)+Nv(\textbf{k}-\textbf{q})$. With this
we obtain a linearized matrix equation: $
[I-\tilde{v}\tilde{G}]\tilde{U}=N\tilde{v}, $ where the sign tilde
is used to identify a matrix. A self-consistent process is
therefore established.

A free electron, or an electron in a perfectly periodic crystal
with effective mass  $m^*$ and wave vector $k$, has its energy
determined by the wave vector, $E(k)=\hbar^2k^2/2m^*$. The
spectral density function (SDF) for a given $E$ in this case is
represented by a delta function in variable $k$, centered at
$k=\sqrt{2m^*E}/\hbar$. On the other extreme, a hidrogen-like
electron bound to an impurity is a well localized particle in the
real space in a region of dimensions given in the scale of the
effective Bohr radius. Its SDF is well spread in the reciprocal
space, having a half-Laurentzian shape  with its maximum at $k=0$
and half width $\Delta k$, in a scale of the inverse of the Bohr
radius. In the past, doped bulk silicon  was studied in a wide
range of concentrations using the Klauder multiple-scattering
technique \cite{ghaz1}. A sharp density of states (DOS) at the
proper impurity energy and a well defined conduction band were
obtained in the extreme diluted regime. A broadening of the
impurity band and a tailing of the conduction band were observed
in the intermediate regime, and a mixing of them appeared in the
highly concentrated regime \cite{ghaz1}, in that case around
$10^{18}$cm$^{-3}$ . This behavior of the DOS comes out of the
disorder due to randomly located scattering centers, leading to a
non-metal-to-metal transition. This transition is associated to
the change in the aspect of the SDF by crossing the Fermi level.

We have performed similar multiple-scattering calculations for
ferromagnetic GaMnAs \cite{usapl} samples having a much higher
density of carriers, of the order of $10^{20}$cm$^{-3}$. For heavy
holes in GaAs, $m^*=0.62m_0$, $m_0$ standing for the electron rest
mass. In the middle of the calculated spin-polarized impurity
bands the SDF has the localized half-Laurentzian-shape. Deep
inside the spin-polarized valence band the SDF has a Laurentzian
shape with a pronounced peak in the neighborhood of a given value
of $k$. However, due to the extremely high concentration of
impurities, the DOS of the impurity and the valence parts of the
DOS collapse, and we obtain two spin-polarized sub-bands. Then, we
need to know about the possible existence of a mobility edge
inside them, i.e., an energy, for a given spin-polarization, where
the character of the hole state changes from localized to
extended, and if this energy lies below the Fermi energy.

Inspired on the arguments above we defined in this work a
parameter $D(n_i,n_s,E,\sigma)$, based on the width and on the
location of the peak of the SDF, as the ratio $
D(n_i,n_s,E,\downarrow)= \frac{k_0}{\Delta_{\downarrow}(E)}$,
where $k_0$ is the wave vector at which the SDF for a given energy
reaches its maximum, and $\Delta_{\downarrow}(E)$ is the width of
the anti-aligned SDF for this energy. Obviously, the higher
$D(n_i,n_s,E,\downarrow)$ the more ``conductive'', or less
localized is the state. For instance, a free hole gives zero width
and a value infinity for the parameter. A single-impurity bound
hole gives a value zero for the parameter. In all cases calculated
the spin-aligned bands were fully localized, therefore they do not
contribute to the transitions. Then, we classify the character of
the samples by a figure of merit defined as
\begin{equation}
D^*(n_i,n_s)=\frac{D(n_i,n_s,E_F,\downarrow)}{\max\{D(n_i,n_s,E_F,\downarrow)\}},
\label{merit}
\end{equation}
where $\max\{D(n_i,n_s,E_F,\downarrow)\}$ refers to the highest
value obtained inside our ensemble, i.e., among the set of $n_i$
and $n_s$ used to characterize  our samples. By obtaining
$D(n_i,n_s,E,\downarrow)$ for energies $E$ going downward from
$E_F$, we can identify a nominal mobility edge $\mu_{\downarrow}$
as the energy for which the figure of merit tends to zero.

\begin{figure}[h]
\includegraphics[width=8.5cm]{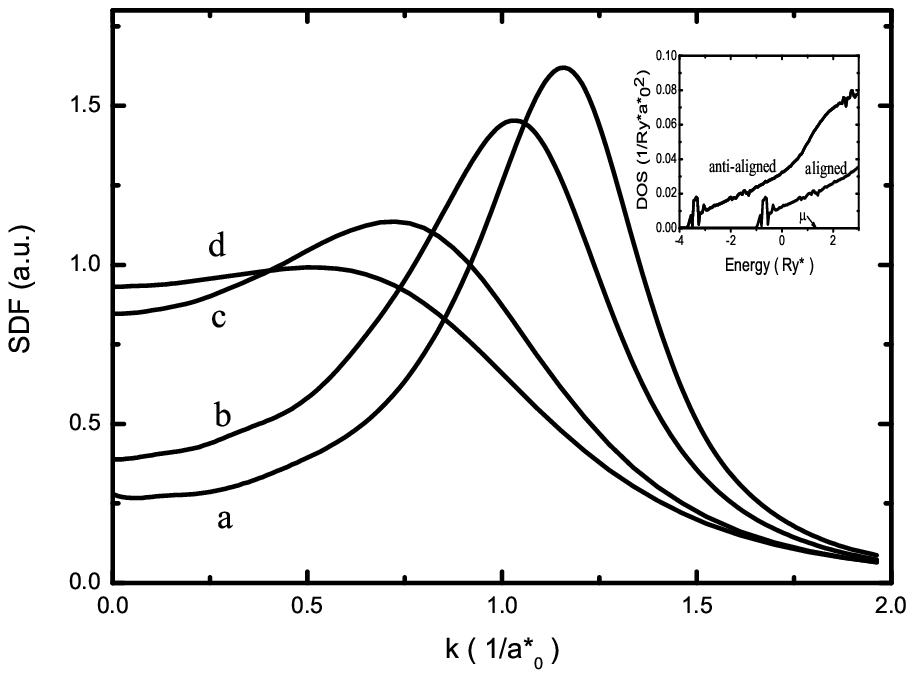}
\caption{Spectral density function (SDF) for $n_i=1.2\times
10^{13}$cm$^{-2}$ and $n_s=6.5\times 10^{12}$cm$^{-2}$. Curves
(a), (b) and (d) correspond to the energies 0.18Ry* above, and
0.43Ry* and 0.38Ry* below the mobility edge, curve (c) at the
nominal mobility edge. The DOS and the nominal mobility edge are
shown in the inset.} \label{fig02}
\end{figure}

A calculation of the SDF for a fully ordered, $\langle M
\rangle=1$, sample with $n_i=1.2\times 10^{13}$cm$^{-2}$ and
$n_s=6.5\times 10^{12}$cm$^{-2}$ is shown in Fig.\ref{fig02}. The
SDF is shown for a few energies below and above the nominal
mobility edge. No sudden change occurs, but the shape of the curve
changes continuously and quickly in the neighborhood of
$\mu=1.38$Ry*. This energy lies in the anti-aligned band in a
region where the impurity band  merged totally into the valence
band, as shown in the inset. This sample is representative of the
best figure of merit obtained in our ensemble.

%
%

\begin{figure}[h]
\includegraphics[angle=-90,width=10cm]{fig04.eps}
\caption{Figure of merit for the metallic character of a sample.
Values are calculated according to Eq.\ref{merit}.} \label{fig03}
\end{figure}

Figure \ref{fig03} shows a 2D plot of the figure of merit in a
wide interval of $n_i$ and $n_s$ with its color scale appended. It
is remarkable the formation of an ``island'' of metallic samples
in the neighborhood of $n_i=1.2\times 10^{13}$cm$^{-2}$ and
$n_s=6.5\times 10^{12}$cm$^{-2}$. Crossing the $n_i$ scale from
left to right we go through the two transitions that have been
observed experimentally. First, there is a non-metal-to-metal
transition, at the far left region. Keeping $n_s$ constant and
increasing $n_i$ we enter a region, starting nearly at
$n_i=8.\times 10^{12}$cm$^{-2}$ and ending at $n_i=1.3\times
10^{13}$cm$^{-2}$, where the figure of merit reaches the maximum
value, shown in the figure by the straight line segment in black.
We leave this island by entering a region where the figure of
merit decreases very quickly, after $n_i=1.6\times
10^{13}$cm$^{-2}$, which is clearly associated to the
metal-to-non-metal transition occurring on the ferromagnetic
samples with high Mn concentration. Representative points with
maximum of the figure of merit correspond to the best quality
sample from the metallic point of view. According to the
conclusions based on the samples shown in Ref. \onlinecite{sado},
they are expected to show the highest Curie temperatures. Another
region where good metallic samples appears is in the upper left
side of our figure. However, this is a region where $n_i$ is low
and $n_s$ high, as compared with $n_i$, and subjected to
oscillations on the indirect exchange potential. It is also
remarkable that the right hand side of the plot shows all states
as localized. This is a region of high disorder, corresponding to
high Mn concentration.

\begin{figure}[h]
\includegraphics[angle=-90,width=10cm]{fig06.eps}
\caption{Effective extended spin-polarized density of charge.}
\label{fig04}
\end{figure}

The values of the density of extended carriers for each metallic
sample, $n^*_s$, are shown in Fig. \ref{fig04}, as the ratio
$n^*_s/n_s$. A Hall measurement is, in fact, a measurement of
$n^*_s$. We can see that, for $n_i=1.2\times 10^{13}$cm$^{-2}$ and
$n_s=6.5\times 10^{12}$cm$^{-2}$ we have roughly 40\% of the total
hole charge as extended. Since this island corresponds to a
fraction $n_s/n_i\approx 0.5$, this means that the ``best
samples''  in our ensemble corresponds to the density of holes
20\% of the Mn concentration, matching very well the experimental
observations.

The entanglement of disorder and spin polarization leads to the
existence of a region in which a non-metal-to-metal transition
occurs, as the Mn concentration is increased. Following a track in
which a favorable relation exists between the Mn and the hole
concentrations, we cross an island of metallic samples and reach
another transition, this time into a non-metallic state.  The
transitions are determined solely by the relative positions of the
Fermi level and the mobility edge. This calculation brings light
to the origin of the magnetic and transport properties of GaMnAs.
At low Mn concentrations the magnetic order is assumed to be
provided by localized hole states, the Curie temperature being, in
consequence, rather low. By increasing the Mn concentration the
average magnetization determined by those localized states
mechanism persists, but as soon as we go through a
non-metal-to-metal transition the extended states come into play.
They now contribute to increase the magnetic order through an
indirect exchange, the Curie temperature raising as we choose a
``good track'' to reach the region of the best samples in the
metallic island. Somewhere in this region the Curie temperature
reaches its maximum value, controlled by the dependence of the
exchange potential on the parameters $n_i$ and $n_s$. By keeping
the Mn concentration increasing and following the track, we leave
the best samples region and we go through the metal-to-non-metal
transition. This is the end of the indirect exchange mediated by
the extended carriers, and the Curie temperature decreases.

This work, supported by CNPq, CAPES FAPERJ, and FAPEMIG, Brazil,
is dedicated to the memory of A. Ghazali.

\end{document}